\documentclass[twocolumn,prd,superscriptaddress,showpacs,floatfix,%
preprintnumbers, nofootinbib,hyperref]{revtex4}  
\usepackage{epsfig}   
\usepackage{bm}  
\usepackage{color}

\newcommand{\be}{\begin{eqnarray}}  
\newcommand{\ee}{\end{eqnarray}}

\begin{document}  

\title{Neutrino mass hierarchy and three-flavor spectral~splits of supernova neutrinos}  
  
\author{Basudeb Dasgupta}
\affiliation{Max-Planck-Institut f\"ur Physik   
(Werner-Heisenberg-Institut), F\"ohringer Ring 6, 80805 M\"unchen, Germany}

\author{Alessandro Mirizzi} 
\affiliation{II Institut f\"ur Theoretische Physik, Universit\"at Hamburg, Luruper Chaussee 149, 22761 Hamburg, Germany}

\author{Irene Tamborra} 
\affiliation{Max-Planck-Institut f\"ur Physik   
(Werner-Heisenberg-Institut), F\"ohringer Ring 6, 80805 M\"unchen, Germany}
\affiliation{Dipartimento Interateneo di Fisica ``Michelangelo Merlin'',
 Via Amendola 173, 70126 Bari, Italy}
\affiliation{INFN, Sezione di Bari,
Via Orabona 4, 70126 Bari, Italy}

\author{Ricard Tom{\`a}s}   
\affiliation{II Institut f\"ur Theoretische Physik, Universit\"at Hamburg, Luruper Chaussee 149, 22761 Hamburg, Germany}

\date{\today}   
   
\preprint{MPP-2010-23}

\begin{abstract}   
It was recently realized that three-flavor effects  could peculiarly modify the development of 
spectral splits induced by collective oscillations, for supernova
neutrinos emitted during the cooling phase of a protoneutron star. 
 We systematically explore this case, explaining how 
the impact of these  three-flavor  effects  depends
on the ordering of the neutrino masses. In inverted mass hierarchy,  
 the solar mass splitting  gives rise to instabilities in regions of 
the (anti)neutrino energy spectra that were otherwise stable 
under the leading two-flavor evolution governed by 
the atmospheric mass splitting and by the 1-3 mixing angle.
 As a consequence, the high-energy spectral splits found in the 
electron (anti)neutrino spectra disappear, and are transferred to other 
flavors. Imperfect adiabaticity leads to smearing 
of spectral swap features. In normal mass hierarchy, the three-flavor 
and the two-flavor instabilities act in the same region of the 
neutrino energy spectrum, leading to only minor departures 
from the two-flavor treatment.
\end{abstract}   
   
   
\maketitle

\section{Introduction}         \label{intro}  
 
The neutrino flux from a core-collapse supernova (SN) is a 
powerful tool to probe fundamental neutrino properties as well as
the dynamics of the explosion~\cite{Raffelt:2007nv,Dighe:2008dq}. 
The  diagnostic role played by neutrinos during a stellar collapse is largely 
associated with the signatures imprinted on the observable 
SN neutrino burst by flavor conversions occurring deep 
inside the star.
 
It has been understood that the paradigm of neutrino 
flavor transformation in supernovae~\cite{Dighe}, based primarily on the 
Mikheyev-Smirnov-Wolfenstein (MSW) effect with the ordinary matter~\cite{Matt}, 
was incomplete. New surprising and unexpected effects have been
found to be important in the region close to the neutrinosphere 
(see \cite{Dighe:2009nr,Duan:2010bg} for recent reviews).
Here the neutrino density is so high that the
neutrino-neutrino interactions dominate the flavor evolution, 
producing collective oscillations. The most important 
observational consequence of $\nu$--$\nu$ interactions is a 
swap of the $\nu_e$ and ${\bar \nu}_e$ spectra with the non-electron 
$\nu_x$ and ${\bar\nu}_x$  spectra in certain energy ranges~\cite{Duan:2006an,Hannestad:2006nj}.

The development  of the spectral swaps is strongly dependent
on the original SN neutrino fluxes. 
In the accretion phase, i.e. $t \lesssim 0.5 $~ms after 
the core-bounce, 
neutrino number fluxes are expected to be ordered as 
$\Phi^0_{\nu_e} > \Phi^0_{{\bar \nu}_e} \gg  \Phi^0_{{\bar \nu}_x}$~\cite{livermore, garching, Keil:2002in}.
In such a scenario, one finds that for normal neutrino mass hierarchy 
(NH,  $\Delta m^2_{\rm atm} = m_3^2-m_{1,2}^2>0$) collective oscillations 
do not play a significant role. 
For inverted hierarchy (IH, $\Delta m^2_{\rm atm}<0$), the end 
of collective oscillations is marked by a complete exchange of the 
$e$ and $x$ flavors for almost all antineutrinos. For neutrinos, the 
exchange happens only above a characteristic energy fixed by the
lepton number conservation, giving 
rise to a \emph{spectral~split}  in their energy 
distributions~\cite{Raffelt:2007cb,Duan:2007fw,Fogli:2007bk,Fogli:2008pt}.

The neutrino number fluxes may be significantly different at 
later times, i.e. during the cooling phase. In Garching 
simulations~\cite{Keil:2002in}, one finds a cross-over among the 
different $\nu$ spectra. As a consequence, the original fluxes exhibit 
a different ordering 
$\Phi^0_{\nu_x} \gtrsim \Phi^0_{{\nu}_e} \gtrsim \Phi^0_{{\bar \nu}_e}$.
A study of this latter case was performed~\cite{Dasgupta:2009mg}, finding
the occurrence of unexpected  \emph{multiple spectral splits} 
for both neutrinos and antineutrinos, in  normal and inverted
mass hierarchies. The rich phenomenology of the spectral splits, and 
its dependence on the original neutrino energy spectra was 
further explored in the extensive study performed 
in~\cite{Fogli:2009rd}.

Most of the collective flavor dynamics of SN neutrinos can be explained in an 
effective two-flavor (2$\nu$) framework~\cite{Dasgupta:2007ws}. Collective oscillations 
are triggered by an instability in the two-flavor ``$H$-sector'' associated 
with the atmospheric mass-squared difference $\Delta m^2_{\rm atm}$
and the mixing angle $\theta_{13}$. Three-flavor (3$\nu$) effects are  due to the 
``$L$-sector,'' governed by the
smaller solar mass splitting $\Delta m^2_{\rm sol} = m_2^2-m_1^2>0$ 
and the mixing angle $\theta_{12}$. These effects have been studied for 
neutrino fluxes typical of the accretion phase~\cite{Dasgupta:2007ws,EstebanPretel:2007yq,Fogli:2008fj,Gava:2008rp}. It was 
recently shown \cite{Dasgupta:2010ae} that they are able to trigger 
collective flavor conversions, even if the mixing angle $\theta_{13}$ 
is exactly zero. 
However, apart from this initial kick,  no new sizeable effect was 
found in the subsequent neutrino flavor evolution. 

Recently, the 
3$\nu$ case  was studied for a scenario relevant to 
the cooling phase~\cite{Friedland:2010sc}. It was found that in the inverted mass 
hierarchy, the presence of the solar sector can ``erase'' the high-energy
spectral splits  that would have occurred in the $\nu_e$ and $\bar{\nu}_e$ 
spectra for a $2\nu$ flavor evolution governed by the 
$\Delta m^2_{\rm atm}$ and $\theta_{13}$. Moreover, the final electron 
antineutrino energy spectrum exhibits a ``mixed'' nature, i.e. the spectral 
swap is not complete. The phenomenological importance of this result is 
underlined by the fact that the high-energy spectral features are expected to be significantly easier to observe at neutrino detectors.
 In~\cite{Friedland:2010sc} these three-flavor effects are associated with an instability in the 
$L$ sector and to the subsequent non-adiabatic flavor evolution driven by $\Delta m^2_{\rm sol}$. Building on this
insight, we   find it worthwhile to take a closer look at the
three-flavor effects during the cooling phase, and to  
understand the origin and 
nature of the 3$\nu$ instability. 

We  explain how the 3$\nu$ effects are crucially dependent on the neutrino mass  hierarchy. 
For normal mass hierarchy, 
there is no fundamental difference between $H$ and $L$ sectors, since in this case both mass splittings are positive and the effective in-medium mixing angles are both small.
Thus, even in a 2$\nu$ set-up one would expect spectral splits driven by the 
$L$-sector parameters, and expect them to be exactly where the splits appear in 
the case of $H$-sector in normal hierarchy. However we find that, it is 
not the case. In fact with typical SN parameters, collective oscillations 
fail to produce any significant conversions for the $L$-sector. This 
is so, because the $L$-sector has a lower natural frequency ($\omega_L=\Delta m^2_{\rm sol}/2E$, where $E$ is a typical SN $\nu$ energy)
 than the $H$-sector and the 
collective interaction strength drops at a rate much faster than it. This 
 does not leave enough time for the instability to grow and makes the 
 collective evolution non-adiabatic. Therefore, spectral splits fail to develop. 
Nevertheless, the system is in an unstable situation. Indeed, 
as we will show, a small perturbation in the initial conditions is enough to develop the collective flavor conversions, producing 
spectral swaps also in this two-flavor case.

In a realistic situation, this initial perturbation for the 
$L$-sector is provided by 3$\nu$ effects that couple this sector to 
the $H$-sector. 
Oscillations in the $H$-sector are communicated to the $L$-sector, 
allowing the instability to grow much faster. The spectral swapping 
still remains less adiabatic than in the $H$-sector.
As a result, for normal mass hierarchy - where both $H$ and $L$-sectors
trigger the same instability and 
``compete'' to convert the high-energy $\nu$ and $\bar{\nu}$ spectra - 
the $H$-sector wins. Three-flavor effects do not cause a significant change.
The interesting bit happens for inverted mass hierarchy. In this case, 
the $H$-sector and the $L$-sector have instabilities in different parts of 
the spectrum and therefore do not compete with each other. Instead, they 
``cooperate'' and act on complementary parts of the energy spectra.
The $L$-sector instability catalyzed by the $H$-sector, operates in the 
high-energy region without hindrance, and causes an additional swap, that 
erases the spectral split found in the $2\nu$ flavor evolution. 
The low adiabaticity in the $L$-sector is responsible for somewhat smeared splits, 
and the effect is particularly important for splits at higher energies, especially
for the antineutrinos. 
 In the remainder of this paper, we illustrate 
these aspects using simple examples and provide a semi-analytical treatment.

The plan of our work is as follows. In Section~2, we present our formalism for the SN neutrino flavor evolution and set up our numerical calculations, i.e. state our inputs for neutrino masses, mixing parameters,
 original supernova neutrino energy spectra and luminosities at late times. In Section~3, we present the 2$\nu$ results
in the $H$ and $L$ sectors, in particular showing the lack of adiabaticity in the $L$-sector and the role of the small perturbations in the initial conditions to circumvent that. In Section~4, we present a complete 3$\nu$ calculation, showing that the $H$-sector catalyzes the $L$-sector and then instabilities in both sectors develop - in cooperation for inverted mass hierarchy, and in competition for normal mass hierarchy. We provide an estimates of the relative adiabaticity of the low energy and high energy splits - explaining the mixed spectra observed in the antineutrino sector. Finally in Section~5, we comment on our results and conclude.

\section{Equations of motion and numerical framework}

Mixed neutrinos are described by matrices of density
$\rho_{\bf p}$ and ${\bar \rho}_{\bf p}$ for each (anti)neutrino mode. The diagonal
entries are the usual occupation numbers whereas the off-diagonal
terms encode phase information. The equations
of motion (EoMs) are~\cite{Sigl:1992fn,Hannestad:2006nj}
\begin{equation}
\textrm{i}\partial_t \rho_{\bf p} = [{\sf H}_{\bf p}, \rho_{\bf p}] \,\ ,
\end{equation}
where the Hamiltonian is 
\begin{equation}
{\sf H}_{\bf p} = \Omega_{\bf p} + V +\sqrt{2} G_F\int \frac{\textrm{d}^3 {\bf q}}{(2\pi)^3}(\rho_{\bf q}-{\bar\rho}_{\bf q})
(1-{\bf v}_{\bf q}\cdot {\bf v}_{\bf q}) \,\ ,
\end{equation}
${\bf v}_{\bf p}$ being the velocity.
The matrix of the vacuum oscillation frequencies for neutrinos is ${\Omega}_{\bf p}=(m_1^2,m_2^2,m_3^2)/2|{\bf p}|$ in the mass basis.
For antineutrinos ${\Omega}_{\bf p} \to -{\Omega}_{\bf p}$.
The matter effect due to the background electron density $n_e$ is represented, in the weak-interaction basis, by $V=\sqrt{2}G_F n_e \textrm{diag}(1,0,0)$.  

In spherical symmetry the EoMs can be expressed as a closed set of differential equations along the radial 
direction~\cite{EstebanPretel:2007ec,Dasgupta:2008cu}.
The factor $(1-{\bf v}_{\bf q}\cdot {\bf v}_{\bf q})$ in the Hamiltonian, implies ``multi-angle'' effects for neutrinos
moving on different trajectories~\cite{Duan:2006an}. However, for realistic supernova conditions the modifications are small, allowing for a single-angle
approximation~\cite{Fogli:2008fj,EstebanPretel:2007ec}. We implement this approximation by launching all neutrinos with $45^\circ$ relative to the radial directions~\cite{EstebanPretel:2007ec,EstebanPretel:2007yq}.

For the numerical illustrations, we take the neutrino mass-squared differences  in vacuum 
to be $\Delta m^2_{\rm atm}= 2\times 10^{-3}$~eV$^2$ and $\Delta m^2_{\rm sol}= 8\times 10^{-5}$~eV$^2$, close to their current best-fit values~\cite{GonzalezGarcia:2010er}.  The values of the mixing parameters relevant for SN neutrino flavor conversions
are $\sin^2\theta_{12} \simeq 0.31$ and $\sin^2 \theta_{13} \lesssim 0.04$~\cite{GonzalezGarcia:2010er}.
The matter effect in the region of collective oscillations (up to a few 100 km) can be accounted for by choosing small 
(matter suppressed) mixing angles~\cite{EstebanPretel:2008ni}, which we take
to be  $\tilde\theta_{13}=\tilde\theta_{12}= 10^{-3}$, and considering as effective mass-square differences
in matter  $\Delta {\tilde m}^2_{\rm atm} = \Delta m^2_{\rm atm} \cos \theta_{13} \simeq \Delta m^2_{\rm atm}$
and $\Delta {\tilde m}^2_{\rm sol} = \Delta m^2_{\rm sol} \cos 2\theta_{12} \simeq 0.4 \Delta m^2_{\rm sol}$~\cite{EstebanPretel:2008ni,Duan:2008za}.  We ignore possible subleading CP violating effects~\cite{Gava:2008rp}
 by setting $\delta_{\rm CP}=0$. MSW conversions  typically occur  after collective 
effects have ceased~\cite{Fogli:2008fj,Gava:2009pj}. Their effects then factorize and can be included separately. Therefore, we neglect them in the following.

The content of a given neutrino species $\nu_\alpha$ with energy $E$ at a position $r$ is given by
\begin{equation}
\rho_{\alpha \alpha} (E,r) = \frac{F_{\nu_\alpha}(E,r)}{F(E,r)} \,\ ,
\end{equation}
where $F_{\nu_\alpha}$ is the energy distribution of $\nu_{\alpha}$ and $F$ is the sum of the energy distributions of all flavors,
for neutrinos and antineutrinos respectively.
For the three relevant SN $\nu$ energy distributions at the  
neutrinosphere $F_{\nu_\alpha}^0$, we take  
\begin{equation}  
\label{Ybeta} F_{\nu_\alpha}^0(E)= {\Phi^0_{{\nu_\alpha}}}\,\varphi_{\nu_\alpha}(E)\ ,  
\end{equation}  
where   
$\Phi_{{\nu_\alpha}}^0 =L_{{\nu_\alpha}}/\langle E_{\nu_\alpha}\rangle$ is the number flux, defined in terms of
the neutrino luminosity $L_{{\nu_\alpha}}$ and the neutrino average energy $\langle E_{\nu_\alpha}\rangle$.
 $\varphi_{\nu_\alpha}(E)$ is the normalized neutrino spectrum ($\int dE  
\; \varphi_{\nu_\alpha}=1$), parametrized as~\cite{Keil:2002in}: 
\begin{equation}  
\label{varphi} \varphi_{\nu_\alpha}(E)=\frac{\beta^\beta}{  
\Gamma(\beta)}\frac{E^{\beta-1}}{\langle E_{\nu_\alpha} 
\rangle^{\beta}} e^{-\beta E/\langle  
E_{\nu_\alpha}\rangle}\; ,  
\end{equation}  
where $\beta$  is a spectral parameter, and
 $\Gamma(\beta)$ is the Euler gamma function. 
The values of the parameters are model dependent~\cite{livermore,garching}.
For our numerical illustrations, we choose  
\begin{equation}  
\label{Eave} (\langle E_{\nu_e} \rangle,\;\langle E_{\bar\nu_e}  
\rangle,\;\langle E_{\bar\nu_x}  
\rangle )  
 = (12,\;15,\;18)\ \mathrm{MeV}\ ,  
\end{equation}  
and $\beta=4$, from the admissible parameter ranges~\cite{Keil:2002in}. 
We take ratios of the fluxes at late-times to be~\cite{garching}
\begin{equation}
\Phi^0_{\nu_e}\;:\;\Phi^0_{{\bar{\nu}}_e}\;:\;\Phi^0_{\nu_x} = 0.85\;:\;0.75\;:\;1.0 \,\ ,
\end{equation}
where we have assumed equal $\mu$ and $\tau$ neutrino and antineutrino initial fluxes.

The strength of the neutrino-neutrino interactions can be parametrized as~\cite{EstebanPretel:2007ec} 
\begin{equation}
\mu_0 = \sqrt{2}G_F |F_{\bar\nu_e}^0-F_{\bar\nu_x}^0| \,\ ,
\end{equation}

where the fluxes are taken at the neutrinosphere radius $R=10$~km.
Numerically, we assume $\mu_0 = 7 \times 10^5$~km$^{-1}$.
When formally deriving the single-angle approximation, one explicitly obtains 
that the radial dependence of the neutrino-neutrino
interaction strength can be written as~\cite{Dasgupta:2008cu}
\begin{equation}
\mu(r) = \mu_0 \frac{R^2}{r^2} C_r \,\ ,
\end{equation}
where the $r^{-2}$ scaling comes from the geometrical flux dilution,
and the collinearity factor
\begin{eqnarray}
C_r &=& 4 \left[\frac{1-\sqrt{1-(R/r)^2}}{(R/r)^2}\right]^2 -1  \nonumber \\
& = & \frac{1}{2}\left(\frac{R}{r}\right)^2 \,\ \,\ \,\ \,\ \textrm{for} \,\  r\to\infty \,\ ,
\end{eqnarray}
arises from the $(1-\cos\theta)$ structure of the neutrino-neutrino interaction.
The asymptotic behavior for large $r$ agrees with what one obtains by considering that all neutrinos are launched at $45^{\circ}$ to the 
radial-direction~\cite{EstebanPretel:2007yq}. 
The known decline of the neutrino-neutrino interaction strength, $\mu(r) \sim r^{-4}$ 
for $r\gg R$, is evident.

\section{Multiple spectral splits in a two-flavor scenario} 

\subsection{$2\nu$ $H$-system}
We start our investigation with the flavor evolution in the $H$-sector characterized by $(\Delta m^2_{\rm atm},
\theta_{13})$.  As explained in \cite{Dasgupta:2009mg}, spectral swaps can develop around energies $E=E_c$ of the original neutrino spectra
(except at $E=0$), where spectra of different flavors cross each other, i.e. at the \emph{crossing points} 
\begin{eqnarray}
F_{\nu_e}(E_c)-F_{\nu_x}(E_c)&=&0 \;\ ,  \nonumber \\
F_{\bar\nu_e}(E_c)-F_{\bar\nu_x}(E_c)&=&0  \;\ , 
\end{eqnarray}
for neutrinos and antineutrinos, respectively.
A given crossing point is \emph{unstable} 
if 
\begin{eqnarray}
d(F_{\nu_e}-F_{\nu_x})/dE<0\quad{\rm for\;IH}\;, \nonumber \\
d(F_{\nu_e}-F_{\nu_x})/dE>0\quad{\rm for\;NH}\;,
\end{eqnarray}
and analogously for antineutrinos.

In Fig.~1, we show the results of the flavor evolution for neutrinos (left panel)
and antineutrinos (right panel) in inverted  mass hierarchy.
In the upper panels we show the initial and final $\nu_e$ and $\nu_x$ energy spectra,
while in the lower panels we show the electron neutrino survival probability $P_{ee}$. 
 Following the instability conditions stated above, one  finds that in IH the spectral swap would develop for neutrinos
 around $E_c\simeq 13$~MeV and for antineutrinos around $E_c\simeq 9$~MeV.
 The development of a spectral swap in the middle of the energy spectra  produces  two splits in the final $\nu$ spectra.
 In particular, in the $\nu_e$ final spectrum the swapped region is between $5$~MeV and $23$~MeV, while for ${\bar\nu}_e$ is between $3$~MeV 
and $17$~MeV.

\begin{figure}[!t]
\begin{center}  
\includegraphics[width=0.95\columnwidth]{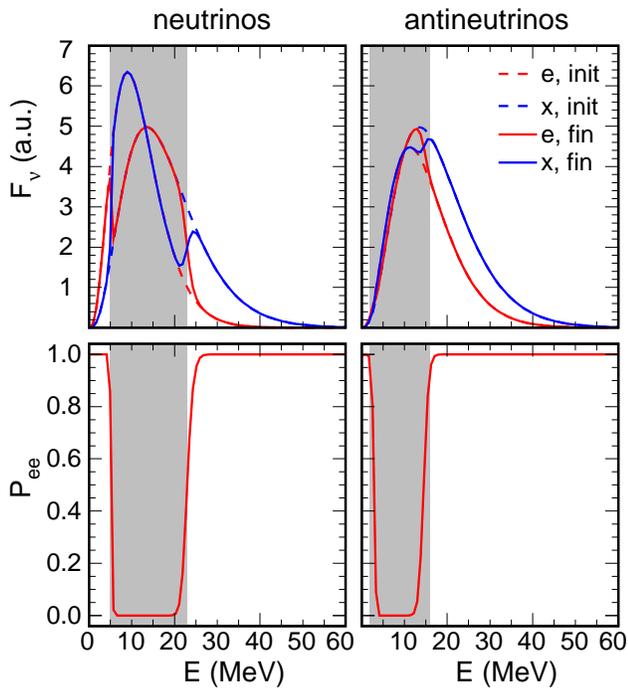}  
\end{center}  
\caption{Flavor evolution in $2\nu$ $H$-sector for neutrinos (left panels) and antineutrinos (right panels)
in inverted mass hierarchy. 
Upper panels: the $\nu_e$ (red) and $\nu_x$ (blue) energy-spectra, before (dashed curves) and after (continuous curves)
the collective oscillations. Lower panels: survival probability $P_{ee}$ for electron (anti)neutrinos.
The grey-regions represent the range in which the spectral swap occurs.
\label{fig:1}}  
\end{figure}  

\begin{figure}[!t]
\begin{center}  
\includegraphics[width=0.95\columnwidth]{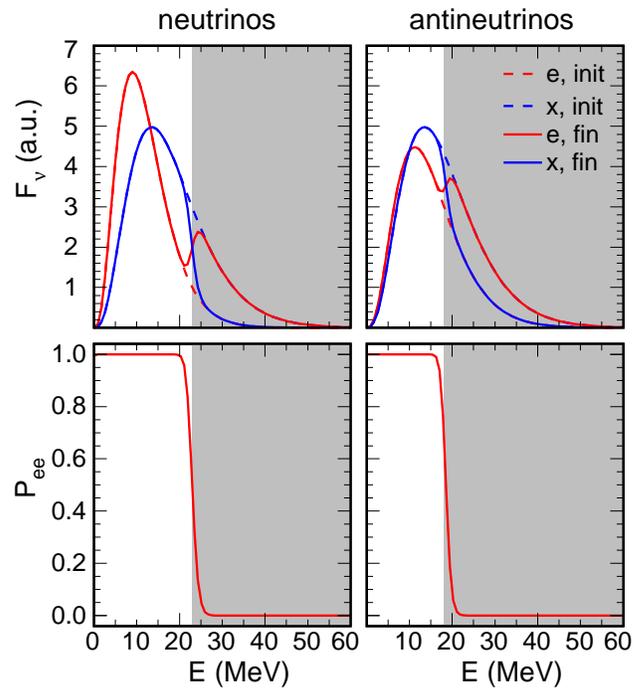}  
\end{center}  
\caption{Flavor evolution in the $2\nu$ $H$-sector for neutrinos (left panels) and antineutrinos (right panels)
in normal mass hierarchy. 
Upper panels: the $\nu_e$ (red) and $\nu_x$ (blue) energy-spectra, before (dashed curves) and after (continuous curves)
the collective oscillations. Lower panels: survival probability $P_{ee}$ for electron (anti)neutrinos.
The grey-regions represent the range in which the spectral swap occurs.
\label{fig:2}}  
\end{figure}  

In Fig.~2, the corresponding results for normal hierarchy are shown. 
In this case the only unstable crossing point is  at $E_c\to \infty$ in the tail of the energy spectra, therefore the resulting $\nu$ and ${\bar\nu}$ spectra exhibit only a high-energy swap and a single split each. In this case, the split is at $E\simeq 23$~MeV for $\nu_e$, and at $E\simeq 17$~MeV for ${\bar\nu}_e$.

From this example, we realize that in the two mass hierarchies the instabilities 
occur around  different and well-separated crossing points. This leads to spectral swaps 
that occur in different energy ranges for the two mass hierarchies. In fact, these ranges are non-overlapping and almost complementary, i.e. the high-energy ends of the swaps in IH are the low-energy ends of the swaps in NH. This implies that if we take the $\nu$ spectra swapped
by the conversions in IH, as an input for conversions in NH, these latter would swap also the high-energy spectrum of the electron (anti)neutrinos, giving the impression that the high-energy split has been ``erased.'' This observation will play an important role in our understanding of the full three-flavor evolution.

\subsection{$2\nu$ $L$-system}

We now consider the $L$-system. Since $\Delta m^2_{\rm sol}>0$,  we expect a behavior similar to that of the $H$-system in normal hierarchy. 
However, we find that in this case no flavor conversion occurs.
This has to be  attributed to two reasons - insufficient growth of the instability, and lack of adiabaticity. The strength of the instability is given by the off-diagonal components in the density matrix. For a simple pendular system with energy $E$, the time-period for growth of off-diagonal components is a 
few times the pendular time-period $\tau_{\rm pend}$~\cite{Hannestad:2006nj,Dasgupta:2009mg}
\begin{equation}
\tau_{\rm pend}\approx\sqrt\frac{2E}{\Delta {\tilde m}^2_{\rm sol} \mu}\; ,
\end{equation} 
which scales logarithmically with the small in-medium mixing angle. 
This time period is roughly  $\sqrt{\Delta {\tilde m}^2_{\rm atm}/\Delta {\tilde m}^2_{\rm sol}} \approx 8$ times larger for the $L$-sector, causing the instability to develop relatively slowly. The slow growth is further exacerbated by a relatively fast decrease in collective neutrino interaction strength $\mu$. 

During the spectral swapping phenomenon, the spectrum near the crossing acts like an inverted pendulum~\cite{Dasgupta:2009mg}. The swap sweeps through the spectrum on each side of the crossing, and the modes at the edge of the swap precess at an average  oscillation frequency ${\kappa}$~\cite{Dasgupta:2009mg}. Adiabaticity requires~\cite{Friedland:2010sc}
\begin{equation}
\left|\frac{d\ln \mu}{dr}\right| < {\kappa} .
\label{eq:adia}
\end{equation}
Since the collective neutrino interaction strength $\mu$ goes as   $r^{-4}$, the rate at which neutrino refractive effects are decreasing is $|d \ln \mu/dr| \simeq 1/50$~km$^{-1}$ at $r\simeq 200$~km, approximately where the swapping takes place. One  finds 
${\kappa} \approx\Delta {\tilde m}^2_{\rm sol}/2 E\simeq 1/400$ km$^{-1}$ 
for a typical energy $ E  \simeq 32$~MeV in the region of the swap, so the adiabaticity condition is not met.

Nevertheless, the $L$-system has the same instability as the $H$-system in NH, around the crossing point $E_c\to \infty$ of the (anti)neutrino energy spectra.
Indeed, if  we slightly perturb the initial conditions of our $\nu$ ensemble, e.g. put explicit off-diagonal terms in the initial neutrino density
 matrix, the instability grows easily and leads to significant flavor conversions.%
\footnote{
The speed-up of flavor instabilities under the effects of very
small seeds in the initial conditions  was already pointed out in~\cite{Sawyer:2004ai}.
}

\begin{figure}[!t]
\begin{center}  
\includegraphics[width=0.95\columnwidth]{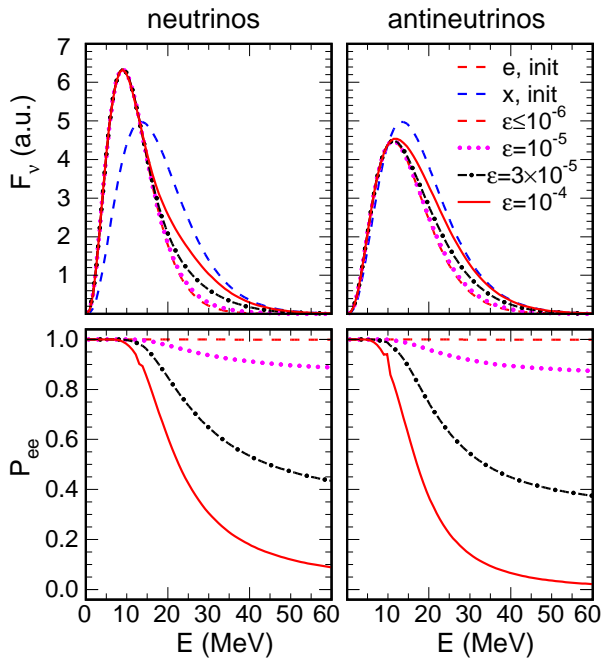}  
\end{center}  
\caption{Flavor evolution in the  $2\nu$ $L$-sector for neutrinos (left panels) and antineutrinos (right panels). 
Upper panels: initial $\nu_e$ (red) and $\nu_x$ (blue) energy-spectra  and
the $\nu_e$ spectrum after the collective oscillations, for different values of an initial off-diagonal component
of the density matrix
$ |\rho_{ex}^0|=\epsilon \times (\rho_{ee}^0 + \rho_{xx}^0)$
(see the text for details). Lower panels: survival probability $P_{ee}$ for electron (anti)neutrinos.
\label{fig:3}}  
\end{figure}  

In Fig.~3,  we show  the effects of these artificially triggered flavor conversions on the electron neutrinos
(left panels) and antineutrinos (right panels). Upper panels show the energy spectrum, while the lower panels display the neutrino survival probability. We find that it is sufficient to take off-diagonal seeds in the density matrix 
 \begin{equation}
 |\rho_{ex}^0|=\epsilon \times (\rho_{ee}^0 + \rho_{xx}^0)\ ,
\end{equation} 
with $\epsilon \gtrsim 10^{-5}$, to produce significant flavor conversions. However, relative to the $H$-sector in the normal hierarchy, the swap is less sharp. This is naturally expected because the adiabaticity does not change significantly by taking an initial perturbation.

\begin{figure}[!t]
\begin{center}  
\includegraphics[width=\columnwidth, height=2.0in]{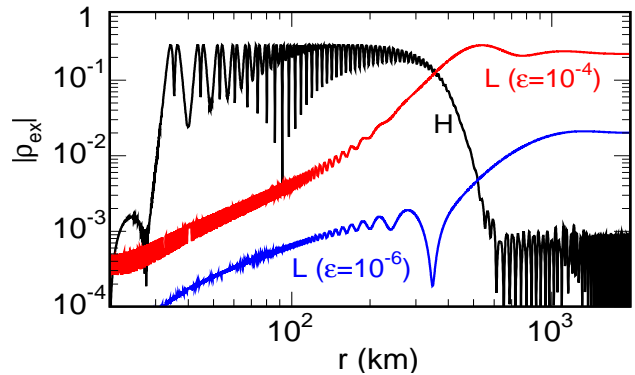}  
\end{center}  
\caption{Growth of the off-diagonal component $|\rho_{ex}|$ of the neutrino density matrix for the $2\nu$ $H$-system
in normal hierarchy and for the $L$-system taking two different values of $|\rho_{ex}^0|=\epsilon \times (\rho_{ee}^0+\rho_{xx}^0)$.  
The neutrino energy is taken to be $E=32$~MeV.
\label{fig:4}}  
\end{figure}  
In Fig.~4, we show the $|\rho_{ex}|$ component of the $\nu$ density matrix for $E= 32$~MeV for three cases: (i) the $H$-system with NH, (ii) the  $L$-system
with $\epsilon=10^{-6}$, and (iii) the  $L$-system seeded with  $\epsilon=10^{-4}$. We realize that the growth of the off-diagonal component
 is almost absent for the $L$-system with $\epsilon=10^{-6}$. Even when there are adequately large off-diagonal elements, as in the
 case of  $\epsilon=10^{-4}$,
 the growth is relatively slow compared to the $H$-system, so that significant flavor changes can start only at large radius ( $r\gtrsim 300$~km). Consequently, due to the violation of adiabaticity, collective flavor conversions do not have enough time to  develop complete splits before the effects of the neutrino-neutrino interactions become negligible.

\section{Multiple spectral splits in a three-flavor scenario} 

\subsection{Speed-up of the $\Delta m^2_{\rm sol}$-instability}

Equipped with the insights of the previous section, we are ready to analyze the
behavior of the flavor conversions in the complete three-flavor scenario.
We work in the rotated basis $(\nu_e,\nu_x,\nu_y) = R^T(\theta_{23})
(\nu_e,\nu_\mu,\nu_\tau)$~\cite{Dasgupta:2007ws}. This is equivalent to take $\theta_{23}=0$ in the neutrino mixing matrix (which makes no difference to $\nu_e$ and $\bar\nu_e$ evolution, if $\mu$ and $\tau$ flavors are treated identically). 
The vacuum Hamiltonian is written as~\cite{Dasgupta:2007ws}
\begin{eqnarray}
\label{Hvac}
\Omega(E) &=& 
\frac{\Delta {\tilde m}^2_{\rm atm}}{2E}  \left(
\begin{array}{ccc}
s^2_{13} & 0 & c_{13}s_{13}\\
0 & 0 & 0 \\
c_{13} s_{13} & 0 & c^2_{13}
 \end{array}\right)  \\
&+&
 \frac{\Delta {\tilde m}^2_{\rm sol}}{2E}
\left(
\begin{array}{ccc}
 c^2_{13}s^2_{12} & c_{12}c_{13}s_{12} & -c_{13}s^2_{12}s_{13}\\
 c_{12}c_{13}s_{12} & c^2_{12} & -c_{12}s_{12}s_{13} \\
-c_{13}s^2_{12}s_{13} & - c_{12}s_{12}s_{13} & s^2_{12}s^2_{13}
 \end{array}
\right)~,\nonumber
\end{eqnarray}
where $c_{ij}=\cos\tilde\theta_{ij}$ and $s_{ij}=\sin\tilde\theta_{ij}$.
 In the limit of $\theta_{13}=0$, the $\nu_e-\nu_y$ ($H$) and the $\nu_e-\nu_x$ ($L$) sectors are completely decoupled.
 
When the neutrino-neutrino interaction
is strong,    all the density matrix mode $\rho_{\bf p}$ stay pinned to each
other, exhibiting  collective flavor conversions. 
Now we review the factorization of the $H$ sector from other sub-leading contributions, 
as shown in~\cite{Dasgupta:2007ws}. For each energy mode, we rewrite our density matrix  and Hamiltonian as
\begin{eqnarray}
\rho&=&\rho^{(0)}+\rho^{(1)}\, ,\\
{\sf H}&=& {\sf H}^{(0)}+{\sf H}^{(1)}\, ,
\end{eqnarray}
where the superscript ${(0)}$ refers to off-components in the $L$-sector and all diagonal components,
 while ${(1)}$ to all others, namely
 \begin{equation}
 \rho^{(0)} = 
 \left(\begin{array}{ccc}
 \rho_{ee} & \rho_{ex} & 0 \nonumber \\
 {\rho^{\ast}_{ex}} & \rho_{xx} & 0 \nonumber \\
 0 & 0 & \rho_{yy}
 \end{array}\right) \,\ ,
 \end{equation}
and 
 \begin{equation}
 \rho^{(1)} = 
 \left(\begin{array}{ccc}
 0 & 0 & \rho_{ey} \nonumber \\
 0 & 0 & \rho_{xy} \nonumber \\
 \rho^{\ast}_{ey} & \rho^{\ast}_{xy} & 0
 \end{array}\right) \,\ .
 \end{equation}
Analogous expressions hold for ${\sf H}^{(0)}$ and ${\sf H}^{(1)}$.

 Note from Eq.~(\ref{Hvac}) that ${\sf H}^{(0)}$, which contains the $e-x$ block, causes oscillations due to $\Delta m^2_{\rm sol}$, while  ${\sf H}^{(1)}$,
containing the $e-y$ off-diagonal terms, gives $\Delta m^2_{\rm atm}$-driven oscillations.
Putting this decomposition into the equations of motion, one finds~\cite{Dasgupta:2007ws}
\begin{eqnarray}
{\rm i}\dot{\rho}^{(0)} &=&  [{\sf H}^{(0)},\rho^{(0)}]+ [{\sf H}^{(1)},\rho^{(1)}]\ ,\label{eq:eomL1} \\
{\rm i}\dot{\rho}^{(1)} &=&  [{\sf H}^{(1)},\rho^{(0)}]+ [{\sf H}^{(0)},\rho^{(1)}]\;. \label{eq:eomL2} 
\end{eqnarray}
Analogous equations hold for the antineutrinos. Note the interesting structure of the EoMs, that is an outcome of the commutation relations - $\dot{\rho}^{(0)}$ depends only on commutators $[{\sf H}^{(0)},\rho^{(0)}]$ and $[{\sf H}^{(1)},\rho^{(1)}]$, while $\dot{\rho}^{(1)}$ depends only on the cross-terms~\cite{Dasgupta:2007ws}. 

In a pure $2\nu$ $L$-system evolution, ${\sf H}^{(1)}$ and $\rho^{(1)}$ are zero. Thus Eq.~(\ref{eq:eomL1}) has only the first term on r.h.s., and Eq.~(\ref{eq:eomL2}) is irrelevant.  Once off-diagonal components of $\rho^{(0)}$, driven by ${\sf H}^{(0)}$, develop asymmetrically in neutrinos and antineutrinos, the instability grows under the action of ${\sf H}^{(0)}$ (proportional to $\Delta m^2_{\rm sol})$, i.e. the pendular time-scale is $\tau_{\rm pend}\sim 1/\sqrt{\omega_{\rm L}\mu}$. The addition of off-diagonal elements to $\rho^{(0)}$ kick-starts the process.

\begin{figure}[!t]
\begin{center}  
\includegraphics[width=\columnwidth]{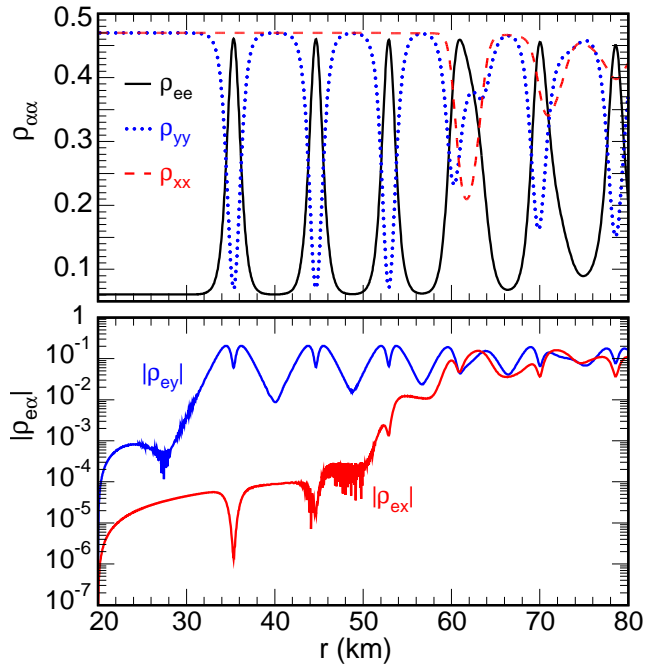}  
\end{center}  
\caption{Initial radial evolution of the diagonal $\rho_{ee}$, $\rho_{xx}$, $\rho_{yy}$ (upper panel) and off-diagonal
$\rho_{ex}$, $\rho_{ey}$ (lower-panel) components of the
neutrino density matrix  for inverted mass hierarchy,  shown for a neutrino energy   $E=32$~MeV.  
\label{fig:5}} 
\end{figure}  

In a complete $3\nu$ system,  ${\sf H}^{(1)}$ is non-zero and thus produces the off-diagonal components in $\rho^{(0)}$ and $\rho^{(1)}$ more quickly. Not only do initial off-diagonal terms get generated in the $L$-sector, but also those terms grow at a much faster rate. The growth is speeded up by the second term in the equation of motion Eq.~(\ref{eq:eomL1}), i.e. by ${\sf H}^{(1)}$ which induces oscillations  $\Delta m^2_{\rm atm}$-dependent at the leading order. 
Therefore, it  leads to a growth of the instability at almost the same rate as for the $H$-system, and much faster than an isolated $L$-system.

 In Fig.~5, we show the diagonal components of the neutrino density matrix $\rho_{ee}$, $\rho_{xx}$, $\rho_{yy}$ (upper panel) 
and the off-diagonal $\rho_{ex}$, $\rho_{ey}$ (lower panel) for a given energy mode with $E= 32$~MeV as a function of $r$ in inverted mass hierarchy. The initial  behavior is qualitatively similar to the one in normal hierarchy (not shown).  
The pendular oscillations of $\rho_{ee}$ begin to develop at $r\simeq35$~km for the $e-y$ sector, and proceed as pure $2\nu$ transitions till $r\simeq 50$~km. Up to this point, $\rho_{ex}$ has not evolved significantly. All off-diagonal components increase rapidly, but the $\rho_{ex}$ starts to develop \emph{only after} $\rho_{ey}$, and saturates at $r\simeq 60$~km, when $e-x$ conversions start. Note that, $\rho_{ex}$ grows faster than a $2\nu$ $L$-system (shown in Fig.~4), as predicted.
The $\theta_{13}$-coupling between the $L$ and $H$ sectors  induces three-flavor effects in the neutrino conversions: The  initial kick, associated with $\Delta m^2_{\rm atm}$,  is necessary to trigger the instability in the $L$-system.

\subsection{Inverted mass hierarchy}

In Fig.~6, we show the complete development of the $\rho_{ee}$, $\rho_{xx}$ and $\rho_{yy}$ components of the density
matrix in inverted mass hierarchy for neutrinos (left panels) and antineutrinos (right panels) for four representative energy modes.
We observe that once the conversions have been started,  the different energy modes for the three neutrino species
oscillate in phase, confirming the collective behavior of the flavor conversions. However, the final fate of $\rho_{ee}$, $\rho_{xx}$ and $\rho_{yy}$
depends on their energy.     
As a result of the three-flavor effects, the  $\nu_e$ mixes with both $\nu_x$ and $\nu_y$. 
Therefore, the $\nu_e$   flavor conversions can be described in terms of
the combined  effects of 
the $L$ and $H$  two-neutrino systems. 
We find that the effects of the  $H$-sector on the $\nu_e-\nu_y$ conversions saturate
before the ones of the $L$-sector, as expected by the hierarchy between the two mass splittings.
As we have discussed in Section~III-A, $\Delta m_{\rm atm}^2<0$ and $\Delta m_{\rm sol}^2>0$  are expected to process complementary parts of the neutrino energy spectra. Therefore, their effects do not interfere in the same energy range.
In particular, in the part of the neutrino energy spectra
unstable under the effects of  $\Delta m_{\rm atm}^2$, the $\rho_{ee}$ and $\rho_{yy}$  would swap, while 
 $\rho_{xx}$, which has been perturbed from its stable equilibrium configuration, would relax to it again.
Conversely, in the part of the neutrino energy spectra
unstable under the effects of  $\Delta m_{\rm sol}^2$, the $\rho_{ee}$ and $\rho_{xx}$   swap, while 
 $\rho_{yy}$ comes back to its initial value.
 
In the three-flavor space, the neutrino ensemble behaves like a pendulum.
Once it is perturbed from its initial configuration, it would evolve toward its stable equilibrium position which may be different for different energy modes. In inverted mass hierarchy, the highest energy modes relax to the $x$ state, while the intermediate ones to the $y$ state,
 while lowest energy modes remain in the $e$ flavor.  
 
\begin{figure}[!t]
\begin{center}  
\includegraphics[width=\columnwidth]{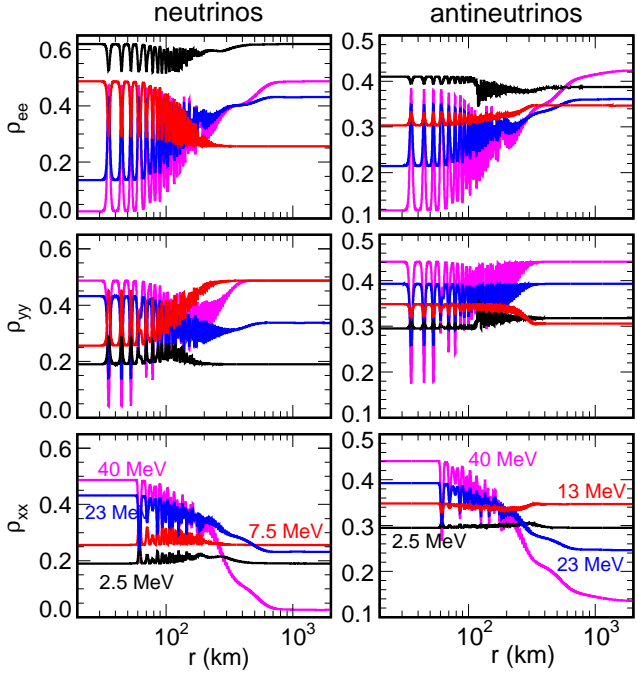}  
\end{center}  
\caption{Three-flavor evolution in inverted mass hierarchy.
Radial evolution of the diagonal components of the density matrix $\rho$ for neutrinos (left panels) and
antineutrinos (right panels)  for  different energy modes.
\label{fig:6}} 
\end{figure}  

\begin{figure}[!t]
\begin{center}  
\includegraphics[width=0.95\columnwidth]{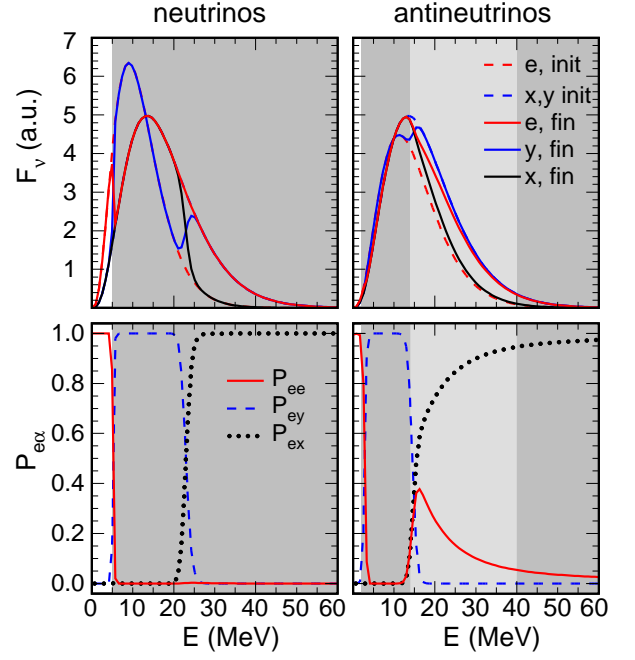}  
\end{center}  
\caption{Three-flavor evolution in inverted mass hierarchy for neutrinos (left panels) and antineutrinos (right panels).
Upper panels: Energy spectra initially (dashed curves) and after collective oscillations (solid curves) for $\nu_e$ (red), $\nu_x$ (black) and $\nu_y$  (blue).
Lower panels: probabilities $P_{ee}$ (solid red curve), $P_{ey}$ (dashed blue curve), $P_{ey}$ (dotted black curve).
The grey bands represent the region where the spectral swaps occur. For antineutrinos the light grey band indicates the region where 
the spectral swap is partial (see text for details). 
\label{fig:7}}  
\end{figure}  
 
For the modes represented in Fig.~6, we see that for $E=2.5$~MeV none of the three neutrino species 
is  affected by significant flavor changes. For $E=7.5$~MeV   the flavor conversions 
 produce a swap between $\rho_{ee}$ and $\rho_{yy}$, while $\rho_{xx}$ comes back to its original value. 
 At  $23$~MeV, we are in the transition region between $e-y$ and $e-x$ conversions, therefore 
  $\rho_{yy}$ and 
$\rho_{xx}$ are both partially swapped into $\rho_{ee}$. 
Finally, at $E=40$~MeV, $\rho_{ee}$ and $\rho_{xx}$ exchange their initial values, while $\rho_{yy}$ returns to its original value. 
Therefore, the behavior of the diagonal components of the density matrix at different energies would produce a single split in the 
$\nu_e$ energy spectrum, since all the $\nu_e$ modes at sufficiently high energy would swap with either $\nu_y$ or  $\nu_x$. Therefore, the high-energy spectral split at $E\simeq 23$~MeV observed in inverted mass hierarchy in the $2\nu$ evolution  of
 Sec.~III-A is washed-out by the three-flavor effects. 
In the antineutrinos (right panels), we find an analogous behavior.
At $E=2.5$~MeV $\rho_{ee}$, $\rho_{xx}$ and $\rho_{yy}$ are essentially unchanged. At $E=13$~MeV
the flavor conversions 
 produce a swap between $\rho_{ee}$ and $\rho_{yy}$, while $\rho_{xx}$ comes back to its original value. 
 For $E=23$~MeV, $\rho_{yy}$ returns to its original value, while 
$\rho_{ee}$ and $\rho_{xx}$ tend to exchange their initial values.
However, as we will discuss later, due to the the less adiabatic behavior  around the  splitting region,
the ${\bar\nu}_e$-${\bar\nu}_x$ swap is not complete.
 Finally, for $E=40$~MeV $\rho_{ee}$ and $\rho_{xx}$ completely convert into each other and $\rho_{yy}$ is stable.

In the upper panels of  Fig.~7, we show the neutrino (left panels) and antineutrino (right panels) spectra 
before and after the complete $3\nu$ flavor conversions in inverted hierarchy.
 In the lower panel we show the corresponding $P_{ee} = P(\nu_e\to\nu_e)$, 
$P_{ex}= P(\nu_e\to\nu_x)$ and $P_{ey}= P(\nu_e\to\nu_y)$ probabilities, 
which can be defined approximately as the off-diagonal elements of the density matrix are small at the end of the collective evolution. 
Namely, these three probabilities read
\begin{eqnarray}
P({\nu_e \to \nu_\alpha}) &=& \frac{\rho_{\alpha\alpha}^f-\rho_{xx}^0}{\rho_{ee}^0-\rho_{xx}^0} \,\ ,
\end{eqnarray}
where $0$ and $f$ indicate the initial and final values of $\rho$, respectively.
{We first consider the neutrino flavor evolution.
As we have already discussed, the effects of the $L$ and $H$ systems process complementary parts of the neutrino energy
spectra.
Indeed, as we can see from the conversion probabilities, in the energy range where $P_{ex}=1$, we have $P_{ey}=0$ and viceversa. In particular, the solar mass splitting $\Delta m_{\rm sol}^2$ induces an electron survival probability $P_{ee}=0$ at high-energies, 
erasing the $H$-sector high-energy split at $E\simeq 23$~MeV, as was pointed out in~\cite{Friedland:2010sc}.}
Thus, the final electron neutrino spectrum  reads
\begin{eqnarray}
F^{f}_{\nu_e}&=&P_{ee}F^0_{\nu_e} + (1-P_{ee})F^0_{\nu_x} \\
&\simeq&
{\Bigg{\{}} \begin{array}{cc} F^0_{\nu_e} & \,\ \,\ \textrm{for} \,\   E \lesssim 5 \,\  \textrm{MeV} \nonumber \\
 F^0_{\nu_x} & \,\ \,\ \textrm{for} \,\   E \gtrsim 5 \,\  \textrm{MeV} 
\;,\\
\end{array}
\end{eqnarray}
Conversely, the behavior of the $y$ and $x$ spectra is the same as the that of the non-electron species in $2\nu$ flavor evolution in IH and NH, respectively, i.e.
\begin{eqnarray}
F^{f}_{\nu_y}&=& P_{ey} F^0_{\nu_e} + (1-P_{ey})F^0_{\nu_x}\\
&\simeq&
{\Bigg{\{}} \begin{array}{cc} F^0_{\nu_x} & \,\ \,\ \textrm{for} \,\   E \lesssim 5 \,\  \textrm{MeV} \nonumber \\
 F^0_{\nu_e} & \,\ \,\ \textrm{for} \,\   5 \,\ \textrm{MeV} \lesssim E \lesssim 23 \,\  \textrm{MeV} \nonumber \\
 F^0_{\nu_x} & \,\ \,\ \textrm{for} \,\    E \gtrsim 23 \,\  \textrm{MeV} 
\;,\\
\end{array}
\end{eqnarray}
and
\begin{flushleft}
\begin{eqnarray}
F^{f}_{\nu_x}&=& P_{ex} F^0_{\nu_e} + (1-P_{ex})F^0_{\nu_x}\\
&=&
{\Bigg{\{}} \begin{array}{cc} F^0_{\nu_x} & \,\ \,\ \textrm{for} \,\   E \lesssim 23 \,\  \textrm{MeV} \nonumber \\
F^0_{\nu_e} & \,\ \,\ \textrm{for} \,\    E \gtrsim 23 \,\  \textrm{MeV} 
\;.\\
\end{array}
\end{eqnarray}
\end{flushleft}
Indeed, the $\nu_x$ and $\nu_y$ spectra will be affected respectively only by the $L$ or the $H$ sector, since
$x-y$ conversions are strongly suppressed as evident from the vacuum Hamiltonian in Eq.~(\ref{Hvac}).
 One should note that while the high-energy spectral split is no longer
present in the $\nu_e$ final energy spectrum,  $\nu_y$ and $\nu_x$  final spectra still present this high-energy feature. Therefore,
 MSW effects in supernova, vacuum mixing  and Earth effects that occur later can make the high-energy split reappear in the observable electron neutrino spectra at Earth.  

Moving on to the  antineutrinos, we realize that the $e-x$ swap is not sharp,  due to the imperfect  adiabaticity in the $L$ sector. This implies that the swap of ${\bar\nu}_e$ and ${\bar\nu}_x$ spectra at intermediate energies ($14$~MeV $\lesssim$ $E \lesssim 40$~MeV) is not complete and this explains the ``mixed'' ${\bar\nu}_e$ spectrum,
observed also  in the numerical simulations in~\cite{Friedland:2010sc}. This effect has to be attributed to the low adiabaticity of the $\Delta m^2_{\rm sol}$-induced conversions, which is particularly visible for antineutrinos.
 The adiabaticity of a spectral split depends on the condition in Eq.~(\ref{eq:adia}).
 The $\bar{\nu}_e$ and $\bar{\nu}_x$ spectra are closer to each other than the corresponding neutrino spectra.
 This implies that in the high-energy region, where the swap takes place,
\begin{equation}
\vert F_{\bar{\nu}_e}(E)-F_{\bar{\nu}_x}(E)\vert\ll\vert F_{{\nu}_e}(E)-F_{{\nu}_x}(E)\vert\; .
\label{eq:fdiff}
\end{equation}
In the final phases of the swapping dynamics, the neutrino and antineutrino spectra evolve quite independently, and the precession frequencies of the two blocks are not governed by a common $\mu$, but by individual $\mu$'s proportional to the flux differences in Eq.~(\ref{eq:fdiff}). They behave essentially as two uncoupled oscillators because the neutrino-neutrino interaction $\mu$ is now smaller than the frequency difference of the two blocks. One can see this clearly in the numerical simulations of Ref.~\cite{Dasgupta:2009mg}, shown at~\cite{movies}. The frequency $\tau_{\rm pend}^{-1}\approx\sqrt{\omega\mu}$ is lower for antineutrinos~\cite{Dasgupta:2009mg}, and thus adiabaticity tends to be broken more severely for antineutrinos.

\subsection{Normal mass hierarchy}

In Fig.~8, we show the complete development of the $\rho_{ee}$, $\rho_{xx}$ and $\rho_{yy}$ components of the density
matrix in normal hierarchy for neutrinos (left panels) and for antineutrinos (right panels) for different energy modes.
For both  neutrinos and antineutrinos, the swapping dynamics occurs mainly between the $\rho_{ee}$ and the $\rho_{yy}$, while $\rho_{xx}$ experiences only nutations around its initial equilibrium value, before relaxing completely
to it when the flavor conversions  are saturated.  
Therefore, the flavor evolution is close to the $2\nu$ $H$-case in normal hierarchy discussed in Sec.~III-A. 
 Indeed,  both
$\Delta m^2_{\rm atm}$ and $\Delta m^2_{\rm sol}$ are positive. Therefore, the $L$-sector would 
behave as a replica of the $H$-sector, but with a smaller mass splitting.
In this condition, both $H$ and $L$ sectors process the same regions of the electron neutrino energy spectra. However, the hierarchy between the two mass splittings produces the dominance of  the $\nu_e-\nu_y$ swaps,  while conversion effects in the $\nu_e-\nu_x$ sector remain  inhibited by adiabaticity violation. The only region where the $L$-instability can compete with $H$ is very close to the split. In particular, for low-energy neutrino modes ($E=2.5,~18$~MeV in  Fig.~8) the $\rho_{ee}$ comes back to its initial value, while at higher energies ($E=23,~40$~MeV) exchanges its initial value
with $\rho_{yy}$. For the antineutrinos, $\rho_{ee}$ remains at its initial value for the modes at $E=2.5$~MeV.
At $E=18$~MeV, we are around  the splitting region where the adiabaticity in the $H$-system is more severely violated, and the 
$e$-$y$ conversions are not complete. Under these conditions, the $L$ instability
can also play a role, producing a weak swap in $\rho_{xx}$. Finally, at $E=40$~MeV, conversions again occur only  between  $e$ and $y$
states, producing a complete swap of the initial $\rho_{ee}$ and  $\rho_{yy}$ values.

\begin{figure}[!t]
\begin{center}  
\includegraphics[width=\columnwidth]{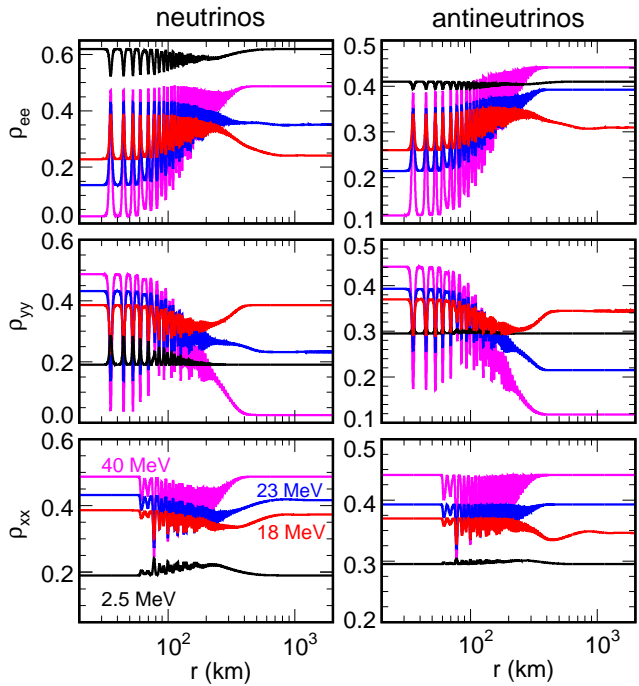}  
\end{center}  
\caption{Three-flavor evolution in normal mass hierarchy.
Radial evolution of the diagonal components of the density matrix $\rho$ for neutrinos (left panels) and
antineutrinos (right panels) for  different energy modes.
\label{fig:8}} 
\end{figure}  

\begin{figure}[!t]
\begin{center}  
\includegraphics[width=0.95\columnwidth]{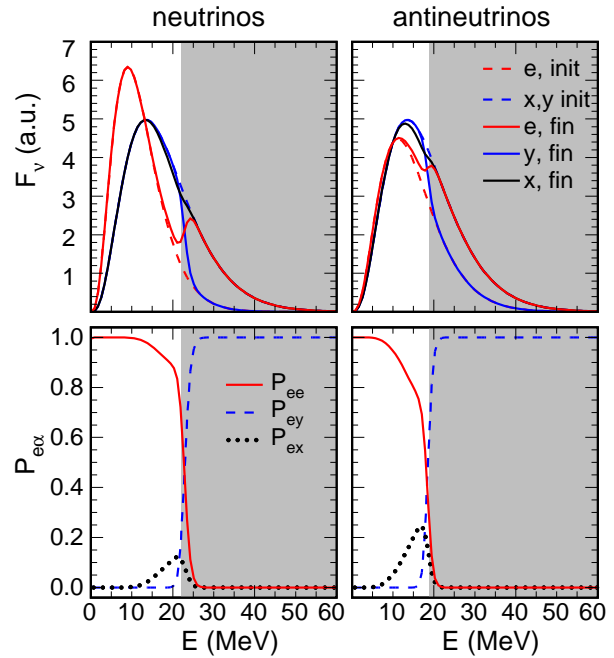}  
\end{center}  
\caption{Three-flavor evolution in normal mass hierarchy for neutrinos (left panels) and antineutrinos (right panels).
Upper panels: 
Energy spectra initially (dashed curves) and after collective oscillations (solid) for $\nu_e$ (red), $\nu_x$ (black) and $\nu_y$ (blue).
Lower panels: corresponding probabilities $P_{ee}$ (solid red curve), $P_{ey}$ (dashed blue curve), $P_{ey}$ (dotted black curve).
The grey bands represent the region where the spectral swaps occur (see text for details). 
\label{fig:9}}  
\end{figure}  
In Fig. 9, we show the neutrinos (left panels) and antineutrino (right panels) spectra before and after the complete $3\nu$ flavor conversions in NH. In the lower panels we represent the corresponding $P_{ee}$, $P_{ex}$ and $P_{ey}$
probabilities. Once more, we realize that the flavor evolution can be mostly described in terms of  two-flavor $\nu_e-\nu_y$ transitions, while the role of $\nu_e-\nu_x$ conversions is subleading.
The spectral splitting features are remarkably similar to the ones found  for the $2\nu$ $H$-system in normal hierarchy
(Sec.~III-A). In particular, the $\nu_e$ spectrum presents a single split at $E\simeq 23$~MeV and $\bar\nu_e$ spectrum around $E\simeq 17$~MeV. 
Only close to the splitting regions, we find subleading effects associated with $\nu_e-\nu_x$ conversions.

\section{Conclusions}

Collective neutrino flavor conversions in supernovae, associated with neutrino-neutrino interactions, have been recognized to induce peculiar spectral swaps among the different neutrino species. 
The development of these features is associated with instabilities in the flavor space. 
In particular, these instabilities would develop around the crossing points of the original SN neutrino spectra. Then, the neutrino mass hierarchy determines if a crossing point is unstable under the effects of the collective oscillations. 
A particularly intriguing case is the one in which
 the original SN neutrino  fluxes exhibit 
an  ordering 
with $\Phi^0_{\nu_x} \gtrsim \Phi^0_{{\nu}_e} \gtrsim \Phi^0_{{\bar \nu}_e}$,  possible 
during the cooling phase. 
In this case, the two-flavor study  realized by~\cite{Dasgupta:2009mg}, found
the occurrence of multiple spectral splits 
for both neutrinos and antineutrinos, depending on the neutrino  
mass hierarchy. 
A recent numerical exploration of this case performed  in~\cite{Friedland:2010sc}, has found that when three-flavor effects are taken into account in inverted mass hierarchy, the high-energy spectral swaps observed in the $2\nu$ evolution 
are erased by effects related to $\Delta m^2_{\rm sol}$.

Motivated by this intriguing result, in our paper we have performed a detailed study of the three-flavor effects in the collective 
oscillations for supernova neutrino spectra typical of the cooling phase. 
We have found that the effects of $\Delta m^2_{\rm sol}$ in the three-flavor evolution are important only in inverted mass hierarchy. 
In this case, the presence of $\Delta m^2_{\rm sol}$ gives rise to instabilities in regions of the neutrino energy spectra
that were stable under the two-flavor evolution governed by $\Delta m^2_{\rm atm}$ and $\theta_{13}$. Therefore, the combinations
of these two different instabilities would produce a wash-out of the high-energy splitting features in the $\nu_e$ and ${\bar\nu}_e$ spectra. 
Conversely, in normal mass hierarchy the three-flavor instabilities and the two-flavor one act in the same regions of the neutrino energy
spectrum, leading only to minor departures from the two-flavor evolution. Essentially, the system behaves like  a pendulum in $3\nu$ flavor space. It can topple towards either the $\nu_y$ state or the $\nu_x$ state. In inverted hierarchy, when the $H$ and $L$ instabilities are in different regions of energy, the pendulum topples towards $\nu_y$ for the $H$-instability, and towards $\nu_x$ for the $L$-instability. In normal hierarchy, when the instabilities are in the same region of energy, the pendulum topples towards $\nu_y$, as the $L$-instability is relatively non-adiabatic.
As a consequence, in inverted mass hierarchy the electron (anti)neutrino spectrum at the end of the collective oscillations would
present only a very low-energy ($E\lesssim 5$~MeV) splitting feature, being completely swapped to the original non-electron spectra 
at higher energies. 

We wish to emphasize that the high-energy splitting features may survive in the observable electron (anti)neutrino spectrum at Earth, even in inverted hierarchy. Indeed, MSW matter effects in SN, vacuum mixing and Earth effects would further mix the $\nu_e$ and $\nu_x$ spectra. The non-electron spectrum at the end of the collective oscillations, still contains high-energy splitting features, since for the non-electron species the collective flavor conversions  have occurred essentially as in the two-flavor case. 
Therefore, especially for neutrinos, which have sharper  spectral swaps, the electron neutrino signal at Earth could
still present observable  splitting features at high energy. Also, due to the lower neutrino luminosity at sufficiently late times, the initial collective interaction strength $\mu_0$ can be somewhat lower than is assumed in~Ref.\cite{Friedland:2010sc} and this work. We found that in certain regions of the spectral parameter space, reducing $\mu_0$ by a factor of 10 makes the three-flavor effect disappear due to a stronger adiabaticity violation in the $L$ sector. In principle, this effect could produce interesting signatures in the time evolution of the SN neutrino signal.

In conclusion, the non-linear equations that govern the flavor evolution of neutrinos emitted during a stellar collapse are a continuous source of surprises and new effects. During this last year, dramatic changes have occurred in the picture consolidated after the initial exploration of collective supernova neutrino oscillations. The discovery of this new three-flavor effect is the most recent of these changes. After our study, it appears that its impact on the  collective neutrino flavor conversions is conceptually and quantitatively well under control.
 
\begin{acknowledgments}  
  We thank Alois Kabelschacht for fruitful discussions during the development of our
 work and for a  careful reading  our manuscript. 
R.T.\ thanks Andreu Esteban-Pretel for useful discussions.
 B.D.\ thanks Alexander Friedland for  useful 
correspondence.
 B.D.\ and I.T.\  were partly supported by the Deutsche
Forschungsgemeinschaft under grant TR-27 ``Neutrinos and Beyond''
and the Cluster of Excellence ``Origin and Structure of the
Universe'' (Munich and Garching). The work of I.T.\ has been partly supported by the Italian
MIUR and INFN through the ``Astroparticle Physics'' research
project.  Her stay
in Munich has been partly supported by a junior fellowship
awarded by the Italian Society of Physics (Borsa SIF
``Antonio Stanghellini'').
\end{acknowledgments}



\begin{thebibliography}{00}   

 
\bibitem{Raffelt:2007nv}  
  G.~G.~Raffelt,  
  ``Supernova neutrino observations: What can we learn?,''  
 [astro-ph/0701677].  
  
\bibitem{Dighe:2008dq}
  A.~Dighe,
  ``Physics potential of future supernova neutrino observations,''
  J.\ Phys.\ Conf.\ Ser.\  {\bf 136}, 022041 (2008)
  [arXiv:0809.2977 [hep-ph]].
   
   
 \bibitem{Dighe}  
  A.~S.~Dighe and A.~Yu.~Smirnov,  
  ``Identifying the neutrino mass spectrum from the neutrino burst from a  
  supernova,''  
  Phys.\ Rev.\ D {\bf 62}, 033007 (2000)  
  [hep-ph/9907423]. 
   
 
  
 \bibitem{Matt}  L.~Wolfenstein,  
				``Neutrino Oscillations In Matter,''  
                Phys.\ Rev.\ D {\bf 17}, 2369 (1978);  
                S. P.~Mikheev and A. Yu.\ Smirnov,  
                ``Resonance Enhancement Of Oscillations In Matter And Solar Neutrino  
				Spectroscopy,''  
                Yad.\ Fiz.\ {\bf 42}, 1441 (1985)  
                [Sov.\ J.\ Nucl.\ Phys.\ {\bf 42}, 913 (1985)].   
                
             
\bibitem{Dighe:2009nr}
  A.~Dighe,
  ``Supernova neutrino oscillations: what do we understand?,''
  J.\ Phys.\ Conf.\ Ser.\  {\bf 203}, 012015 (2010)
  [arXiv:0912.4167 [hep-ph]].
  
\bibitem{Duan:2010bg}
  H.~Duan, G.~M.~Fuller and Y.~Z.~Qian,
  ``Collective Neutrino Oscillations,''
  [arXiv:1001.2799 [hep-ph]].

\bibitem{Duan:2006an} H.~Duan, G.M.~Fuller, J.~Carlson and Y.Z.~Qian,  
    ``Simulation of coherent non-linear neutrino flavor  
    transformation in the supernova environment. I: Correlated  
    neutrino trajectories,'' Phys.\ Rev.\ D {\bf 74}, 105014 (2006)  
    [astro-ph/0606616].  

  
\bibitem{Hannestad:2006nj}  
  S.~Hannestad, G.~G.~Raffelt, G.~Sigl and Y.~Y.~Y.~Wong,  
  ``Self-induced conversion in dense neutrino gases:  
  Pendulum in flavor space,''  
  Phys.\ Rev.\ D {\bf 74}, 105010 (2006);  
  Erratum ibid.\ {\bf 76}, 029901 (2007)  
  [astro-ph/0608695].  

\bibitem{livermore}
T.~Totani, K.~Sato, H.~E.~Dalhed and J.~R.~Wilson,
``Future detection of supernova neutrino burst and explosion mechanism,''
Astrophys.\ J.\  {\bf 496}, 216 (1998)
[astro-ph/9710203].

\bibitem{garching}
R.~Buras, H.~T.~Janka, M.~T.~Keil, G.~G.~Raffelt and M.~Rampp,
``Electron-neutrino pair annihilation: A new source for muon and tau
neutrinos in supernovae,'' 
Astrophys.\ J.\  {\bf 587}, 320 (2003)
[astro-ph/0205006].

    
\bibitem{Keil:2002in}  
  M.~T.~Keil, G.~G.~Raffelt and H.~T.~Janka,  
  ``Monte Carlo study of supernova neutrino spectra formation,''  
  Astrophys.\ J.\  {\bf 590}, 971 (2003)  
  [astro-ph/0208035].  
  
  
  \bibitem{Raffelt:2007cb}  
  G.~G.~Raffelt and A.~Yu.~Smirnov,  
  ``Self-induced spectral splits in supernova neutrino fluxes,''  
  Phys.\ Rev.\  D {\bf 76}, 081301 (2007)  
  [arXiv:0705.1830 [hep-ph]].  
 
  
\bibitem{Duan:2007fw}  
 H.~Duan, G.~M.~Fuller and Y.~Z.~Qian,  
 ``A simple picture for neutrino flavor transformation  
 in supernovae,''  
 Phys.\ Rev.\  D {\bf 76}, 085013 (2007)  
 [arXiv:0706.4293 [astro-ph]].  
  

  
\bibitem{Fogli:2007bk}  
 G.~L.~Fogli, E.~Lisi, A.~Marrone and A.~Mirizzi,  
 ``Collective neutrino flavor transitions in supernovae and the role of  
 trajectory averaging,''  
 JCAP {\bf 0712}, 010 (2007)  
 [arXiv:0707.1998 [hep-ph]].  

\bibitem{Fogli:2008pt}
  G.~L.~Fogli, E.~Lisi, A.~Marrone, A.~Mirizzi and I.~Tamborra,
  ``Low-energy spectral features of supernova (anti)neutrinos in inverted
  hierarchy,''
  Phys.\ Rev.\  D {\bf 78}, 097301 (2008)
  [arXiv:0808.0807 [hep-ph]].


\bibitem{Dasgupta:2009mg}
  B.~Dasgupta, A.~Dighe, G.~G.~Raffelt and A.~Y.~Smirnov,
  ``Multiple Spectral Splits of Supernova Neutrinos,''
  Phys.\ Rev.\ Lett.\  {\bf 103}, 051105 (2009)
  arXiv:0904.3542~[hep-ph].
  
\bibitem{Fogli:2009rd}
  G.~Fogli, E.~Lisi, A.~Marrone and I.~Tamborra,
  ``Supernova neutrinos and antineutrinos: ternary luminosity diagram and
  spectral split patterns,''
  JCAP {\bf 0910}, 002 (2009)
  [arXiv:0907.5115 [hep-ph]].

\bibitem{Dasgupta:2007ws}
  B.~Dasgupta and A.~Dighe,
  ``Collective three-flavor oscillations of supernova neutrinos,''
  Phys.\ Rev.\  D {\bf 77}, 113002 (2008)
  [arXiv:0712.3798 [hep-ph]].

\bibitem{EstebanPretel:2007yq}
  A.~Esteban-Pretel, S.~Pastor, R.~Tomas, G.~G.~Raffelt and G.~Sigl,
  ``Mu-tau neutrino refraction and collective three-flavor transformations in
  supernovae,''
  Phys.\ Rev.\  D {\bf 77}, 065024 (2008)
  [arXiv:0712.1137 [astro-ph]].


\bibitem{Fogli:2008fj}
  G.~Fogli, E.~Lisi, A.~Marrone and I.~Tamborra,
  ``Supernova neutrino three-flavor evolution with dominant collective
  effects,''
  JCAP {\bf 0904}, 030 (2009)
  [arXiv:0812.3031 [hep-ph]].

\bibitem{Gava:2008rp}
  J.~Gava and C.~Volpe,
  ``Collective neutrinos oscillation in matter and CP-violation,''
  Phys.\ Rev.\  D {\bf 78}, 083007 (2008)
  [arXiv:0807.3418 [astro-ph]].
  
\bibitem{Dasgupta:2010ae}
  B.~Dasgupta, G.~G.~Raffelt and I.~Tamborra,
  Phys.\ Rev.\  D {\bf 81}, 073004 (2010)
  [arXiv:1001.5396 [hep-ph]].

\bibitem{Friedland:2010sc}
  A.~Friedland,
  ``Self-refraction of supernova neutrinos: mixed spectra and three-flavor
  instabilities,''
  Phys.\ Rev.\ Lett.\  {\bf 104}, 191102 (2010)
  [arXiv:1001.0996 [hep-ph]].


\bibitem{Sigl:1992fn}
  G.~Sigl and G.~Raffelt,
  ``General kinetic description of relativistic mixed neutrinos,''
  Nucl.\ Phys.\  B {\bf 406}, 423 (1993).

\bibitem{EstebanPretel:2007ec}  
 A.~Esteban-Pretel, S.~Pastor, R.~Tom\`as,  
 G.~G.~Raffelt and G.~Sigl,  
 ``Decoherence in supernova neutrino transformations  
 suppressed by deleptonization,''  
 Phys.\ Rev.\ D {\bf 76}, 125018 (2007)  
 [arXiv:0706.2498 [astro-ph]].  
 
\bibitem{Dasgupta:2008cu}
  B.~Dasgupta, A.~Dighe, A.~Mirizzi and G.~G.~Raffelt,
  ``Collective neutrino oscillations in non-spherical geometry,''
  Phys.\ Rev.\  D {\bf 78}, 033014 (2008)
  [arXiv:0805.3300 [hep-ph]].

\bibitem{GonzalezGarcia:2010er}
  M.~C.~Gonzalez-Garcia, M.~Maltoni and J.~Salvado,
  ``Updated global fit to three neutrino mixing: status of the hints of theta13
  > 0,''
  JHEP {\bf 1004}, 056 (2010)
  [arXiv:1001.4524 [hep-ph]].


\bibitem{EstebanPretel:2008ni}
  A.~Esteban-Pretel, A.~Mirizzi, S.~Pastor, R.~Tomas, G.~G.~Raffelt, P.~D.~Serpico and G.~Sigl,
  ``Role of dense matter in collective supernova neutrino transformations,''
  Phys.\ Rev.\  D {\bf 78}, 085012 (2008)
  arXiv:0807.0659 [astro-ph].

\bibitem{Duan:2008za}
  H.~Duan, G.~M.~Fuller and Y.~Z.~Qian,
  ``Stepwise Spectral Swapping with Three Neutrino Flavors,''
  Phys.\ Rev.\  D {\bf 77}, 085016 (2008)
  [arXiv:0801.1363 [hep-ph]].


\bibitem{Gava:2009pj}
  J.~Gava, J.~Kneller, C.~Volpe and G.~C.~McLaughlin,
  ``A dynamical collective calculation of supernova neutrino signals,''
  Phys.\ Rev.\ Lett.\  {\bf 103}, 071101 (2009)
  [arXiv:0902.0317 [hep-ph]].

\bibitem{Sawyer:2004ai}
  R.~F.~Sawyer,
  ``'Classical' instabilities and 'quantum' speed-up in the evolution of
  neutrino clouds,''
  [hep-ph/0408265].
  
\bibitem{movies}
B.~Dasgupta, A.~Dighe, G.~G.~Raffelt and A.~Y.~Smirnov,
Animated figures available at 
http://www.mppmu.mpg.de/supernova/multisplits.


\end{thebibliography}
\end{document}